\documentclass{ceurart}

\usepackage{listings}
\usepackage{graphicx}
\usepackage{booktabs}
\usepackage{subcaption}
\usepackage{multirow}
\usepackage{amsmath}
\usepackage{tikz}
\usetikzlibrary{arrows.meta, positioning, shapes.geometric, fit}
\usepackage[htt]{hyphenat}
\usepackage{pgfplots}
\usepackage{pgfplotstable}
\pgfplotsset{compat=1.18}
\usepackage{enumitem}
\usepackage{tabularx}
\usepackage{tcolorbox}
\usepackage[shortcuts]{extdash}
\usepackage{graphicx}
\usepackage{subcaption}
\usepackage{float}     
\usepackage{placeins}  
\usepackage{subcaption}
\usepackage{caption}
\usepackage{etoolbox}
\usepackage{eso-pic}
\usepackage{calc}

\AddToShipoutPictureFG{
  \AtPageUpperLeft{
    \put(\LenToUnit{\dimexpr\paperwidth/2-\textwidth/2\relax}, \LenToUnit{-1.5cm}){
      \vbox{
        \centering
        \begin{minipage}{\textwidth}
          {\small\textbf{Published in the Proceedings of the SmaLLEXT Workshop at the CIKM 2025}}\par
          \vspace{2pt}
          \hrule height 0.4pt
        \end{minipage}
      }
    }
  }
}

\begin{document}
\copyrightyear{2025}
\copyrightclause{Copyright for this paper by its authors.
  Use permitted under Creative Commons License Attribution 4.0
  International (CC BY 4.0).}

\conference{Woodstock'22: Symposium on the irreproducible science,
  June 07--11, 2022, Woodstock, NY}
\conference{SmaLLEXT@CIKM'25: The 1st Workshop on Small and Efficient Large Language Models for Knowledge Extraction,\\ Nov 14, 2025, Seoul, Korea}

\title{Exploring Approaches for Detecting Memorization of Recommender System Data in Large Language Models}

\author[1]{Antonio Colacicco}[%
email=a.colacicco1@studenti.poliba.it,
]
\cormark[1]

\author[1]{Vito Guida}[%
email=v.guida@studenti.poliba.it,
]
\cormark[1]

\author[1]{Dario Di Palma}[%
email=dario.dipalma@poliba.it,
]
\fnmark[1]

\author[1]{Fedelucio Narducci}[%
email=fedelucio.narducci@poliba.it,
]

\author[1]{Tommaso Di Noia}[%
email=tommaso.dinoia@poliba.it,
]

\address[1]{Politecnico di Bari, Bari, Italy}

\fntext[1]{Corresponding author.}
\cortext[1]{These authors contributed equally.}

\begin{abstract}
  Large Language Models (LLMs) are increasingly applied in recommendation scenarios due to their strong natural language understanding and generation capabilities. However, they are trained on vast corpora whose contents are not publicly disclosed, raising concerns about data leakage. Recent work has shown that the MovieLens-1M dataset is memorized by both the LLaMA and OpenAI model families, but the extraction of such memorized data has so far relied exclusively on manual prompt engineering.
  
  In this paper, we pose three main questions: \emph{Is it possible to enhance manual prompting? Can LLM memorization be detected through methods beyond manual prompting? And can the detection of data leakage be automated?}
  
  To address these questions, we evaluate three approaches: (i) jailbreak prompt engineering; (ii) unsupervised latent knowledge discovery, probing internal activations via Contrast–Consistent Search (CCS) and Cluster-Norm; and (iii) Automatic Prompt Engineering (APE), which frames prompt discovery as a meta-learning process that iteratively refines candidate instructions.
  
  Experiments on MovieLens-1M using LLaMA models show that jailbreak prompting does not improve the retrieval of memorized items and remains inconsistent; CCS reliably distinguishes genuine from fabricated movie titles but fails on numerical user and rating data; and APE retrieves item-level information with moderate success yet struggles to recover numerical interactions. These findings suggest that automatically optimizing prompts is the most promising strategy for extracting memorized samples.
\end{abstract}

\maketitle

\section{Introduction}
Large Language Models (LLMs) have transformed natural language processing and are increasingly integrated into sentiment analysis \cite{DBLP:conf/acl/PalmaBSANN25}, conversational agents \cite{biancofiore2025conversational}, search engines \cite{DBLP:conf/www/Liu00L0DLN24}, and recommender systems \cite{DBLP:conf/recsys/Palma23}. Their training typically involves ingesting vast amounts of text from diverse sources, yet the specific training data are not publicly disclosed. While such breadth enables strong generalization, it also raises concerns about data leakage.

The task of determining, given a record and a trained model, whether the record was part of the model’s training set is formalized as a Membership Inference Attack (MIA)~\cite{DBLP:conf/sp/ShokriSSS17} and originated as a privacy-risk auditing technique.
In MIAs, two main attack settings are considered: black-box attacks, where the adversary can only query the model and observe its outputs (e.g., predicted labels) without access to internal parameters or gradients, and white-box attacks, where the adversary has full access to the model’s architecture, parameters, and intermediate computations, enabling more direct and often more accurate detection of training-set membership. The canonical black-box attack trains shadow models to distinguish “in” from “out” posterior behaviors \citep{DBLP:conf/sp/ShokriSSS17}, while white-box variants exploit internal signals (e.g., gradients, model updates) and have been extended to federated learning settings \citep{DBLP:conf/sp/NasrSH19}.

Starting from MIA, researchers have begun to address the data leakage problem by defining and quantifying LLM memorization. For example, \citet{DBLP:conf/iclr/CarliniIJLTZ23} found that the GPT-J-6B model memorized at least 1\% of the Pile dataset~\cite{DBLP:journals/corr/abs-2101-00027}, while \citet{DBLP:conf/icse/Al-KaswanID24} were able to extract 56\% of the coding samples used to train GPT-Neo.

In the context of recommender systems, recent work~\cite{DBLP:conf/sigir/PalmaMSANN25} revealed that the MovieLens-1M dataset is memorized within both the LLaMA and OpenAI model families. For example, GPT-4o was able to retrieve 80.76\% of item samples, while LLaMA-3.3 70B retrieved 7.65\%. Moreover, recommender performance was found to correlate with the degree of such memorization. This finding confirms that generative models can store and reproduce training examples when prompted. However, for LLM based recommender systems, this presents an important limitation: MovieLens-1M is a widely used benchmark dataset often evaluated using classical RecSys protocols~\cite{DBLP:conf/sigir/YangMSALCZ24, DBLP:conf/naacl/LyuJZXWZCLTL24, DBLP:conf/iir/PalmaSABNCN24, DBLP:conf/um/PalmaBANN25}, meaning that test sets may already be memorized by the models, thereby undermining the reliability of evaluation results.

This paper investigates whether memorized instances from MovieLens\hyp{1M} can be extracted through multiple strategies, and whether probing methods can detect such memorization. Our work directly builds upon the preliminary study by~\citet{DBLP:conf/sigir/PalmaMSANN25}, which relied solely on manual prompt engineering to test for memorization in LLMs. 

We therefore pose the following research question:
\begin{quote}
\emph{Is it possible to enhance manual prompting? Can LLM memorization be detected through methods beyond manual prompting? And can the detection of data leakage be automated?}
\end{quote}

To address this question, we evaluate three complementary families of techniques:
(i) Jailbreaking prompt engineering (white-box), to assess whether jailbroking help to reveal memorized data;
(ii) Unsupervised latent knowledge discovery (black-box), probing internal activations using Contrast–Consistent Search (CCS)~\cite{DBLP:conf/iclr/BurnsYKS23} and Cluster-Norm~\cite{DBLP:conf/emnlp/LauritoMDYH24}; and
(iii) Automatic Prompt Engineering~\cite{DBLP:conf/iclr/ZhouMHPPCB23} (white-box), formulating prompt generation as a meta-learning process that iteratively refines candidate instructions.

Our contributions are threefold:
\begin{itemize}[itemsep=0pt, parsep=0pt, topsep=0pt, partopsep=0pt, leftmargin=15pt]
    \item We contextualize and extend the findings of~\citet{DBLP:conf/sigir/PalmaMSANN25} by systematically evaluating jailbreaking, unsupervised, and automated methods for detecting memorization of MovieLens-1M.
    \item We conduct a detailed experimental study on public LLaMA-1B and 3B models, quantifying the efficacy of each technique across item, user, and rating fields.
    \item We provide qualitative and quantitative analyses, offering actionable recommendations for practitioners and outlining future research directions.
\end{itemize}
To our knowledge, this is the first comprehensive comparison of manual, unsupervised, and automated probing methods on recommender system data.

\section{Methodology}
\subsection{Data and Preliminaries}
To ensure a fair comparison with previous work and enable the use of white-box methods, we focus on the MovieLens-1M dataset and LLaMA models, allowing a direct evaluation of the proposed methods’ capabilities in detecting memorization. 

The dataset is organized into three files: \texttt{users.dat}, listing user identifiers and demographic attributes; \texttt{movies.dat}, containing movie titles and genres; and \texttt{ratings.dat}, with the user interactions triples. Table~\ref{tab:datasets} summarizes the basic statistics.
\begin{table}[t!]
  \centering
  \caption{Statistics of the MovieLens‑1M dataset.}
  \label{tab:datasets}
  \begin{tabular}{lccc}
    \toprule
    File & Records & Features & Raw Record\\
    \midrule
    \texttt{users.dat} & 6.040 & 5 & userID::gender::age::occupation::zip\\
    \texttt{movies.dat} & 3.952 & 3 & movieID::title::genres\\
    \texttt{ratings.dat} & 1.000.209 & 4 & userID::movieID::rating::timestamp\\
    \bottomrule
  \end{tabular}
\end{table}

In the following, we describe each methodology in detail and explain how it is applied for the dataset discovery.

\subsection{Jailbreak Prompt Engineering}
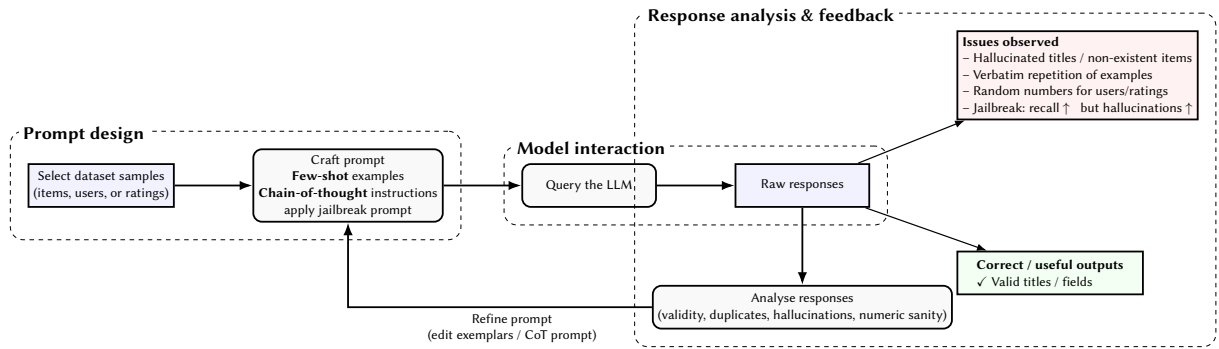
\begin{figure}[t]
\centering
\caption{Jailbreak prompt-engineering workflow: select dataset samples; craft a prompt with few-shot examples and chain-of-thought instructions incorporating jailbreaking prompt; query the LLM; manual analysis of responses for validity, duplication, and hallucinations; then iterate on the prompt.}
\resizebox{\columnwidth}{!}{%
\begin{tikzpicture}[
    font=\sffamily\footnotesize,
    node distance=12mm and 18mm,
    >={Latex},
    process/.style={rectangle, rounded corners=2mm, draw, very thick, align=center, minimum width=31mm, minimum height=10mm, fill=gray!5},
    data/.style={rectangle, draw, very thick, align=center, minimum width=31mm, minimum height=10mm, fill=blue!5},
    result/.style={rectangle, draw, very thick, align=left, minimum width=43mm, minimum height=10mm, fill=green!5},
    issue/.style={rectangle, draw, very thick, align=left, minimum width=48mm, minimum height=10mm, fill=red!5},
    note/.style={rectangle, draw, align=left, rounded corners=1mm, fill=yellow!10},
    groupbox/.style={draw, rounded corners=3mm, inner sep=4mm, dashed}
]

\node[data]    (samples) {Select dataset samples\\(items, users, or ratings)};
\node[process, right=of samples] (craft) {Craft prompt\\\textbf{Few-shot} examples\\\textbf{Chain-of-thought} instructions\\apply jailbreak prompt};
\node[process, right=of craft]   (query) {Query the LLM};
\node[data,    right=of query]   (responses) {Raw responses};

\node[process, below=18mm of responses] (analyze) {Analyse responses\\(validity, duplicates, hallucinations, numeric sanity)};
\node[result,  below right=10mm and 20mm of responses] (good) {\textbf{Correct / useful outputs}\\\(\checkmark\) Valid titles / fields};
\node[issue,   above right=10mm and 20mm of responses] (bad) {\textbf{Issues observed}\\-- Hallucinated titles / non-existent items\\-- Verbatim repetition of examples\\-- Random numbers for users/ratings\\-- Jailbreak: recall $\uparrow$ \; but hallucinations $\uparrow$};


\draw[very thick, ->] (samples) -- (craft);
\draw[very thick, ->] (craft) -- (query);
\draw[very thick, ->] (query) -- (responses);
\draw[very thick, ->] (responses) -- (analyze);
\draw[very thick, ->] (analyze) -| node[pos=0.23, below, align=center]{Refine prompt\\(edit exemplars / CoT prompt)} (craft);

\draw[thick, ->] (responses) -- (good);
\draw[thick, ->] (responses) -- (bad);

\node[groupbox, fit=(samples)(craft)] (g1) {};
\node[groupbox, fit=(query)(responses)] (g2) {};
\node[groupbox, fit=(analyze)(good)(bad)] (g3) {};

\node[anchor=north west, xshift=2mm, yshift=2mm, fill=white, inner sep=1mm] 
    at (g1.north west) {\bfseries\large Prompt design};

\node[anchor=north west, xshift=2mm, yshift=2mm, fill=white, inner sep=1mm] 
    at (g2.north west) {\bfseries\large Model interaction};

\node[anchor=north west, xshift=2mm, yshift=2mm, fill=white, inner sep=1mm] 
    at (g3.north west) {\bfseries\large Response analysis \& feedback};

\end{tikzpicture}
}
\label{fig:manual_workflow}
\end{figure}

Jailbreak prompt engineering involves crafting adversarial prompts that bypass safety filters and alignment mechanisms in LLMs. Early work explored attacks such as prompt injection~\cite{DBLP:conf/ccs/AbdelnabiGMEHF23} and prompt leakage~\cite{DBLP:conf/ccs/0002Y0BC24}, where malicious instructions exploit a model’s instruction\hyp{following} behavior to override content restrictions. Techniques including role-playing scenarios, obfuscation of restricted terms, and multi-turn reasoning traps have been shown to elicit otherwise blocked outputs~\cite{DBLP:conf/ccs/ShenC0SZ24,DBLP:conf/coling/LiWLWDLWZH25}.

While commonly studied for harmful or unsafe content generation, we examine their potential in a controlled setting to probe memorization of the MovieLens-1M dataset. Specifically, we test whether jailbreak prompts can retrieve specific dataset entries. Our iterative workflow (Figure~\ref{fig:manual_workflow}) comprises: (i) Prompt design, embedding raw dataset samples into few-shot templates augmented with jailbreak-style instructions from~\citet{DBLP:journals/corr/abs-2503-05264}, fabricating a conversation history that primes the model for compliance (see appendix Figure~\ref{fig:jailbreak_example} for further details); (ii) Model interaction, issuing the crafted prompt to the Llama-1B; and (iii) Response analysis, manually evaluating outputs for validity, duplication, hallucination, and numeric plausibility.

Prompts are iteratively refined by adjusting exemplars and instructions until achieving a balance between recall and accuracy.

\subsection{Unsupervised Latent Knowledge Discovery}
An emerging research area on the interpretability of LLMs leverages their internal activations to unveil fine-grained properties, including linguistic properties~\cite{DBLP:conf/emnlp/VulicPLGK20}, factual knowledge~\cite{petroni-etal-2019-language, DBLP:conf/acl/ServedioBPAN25}, and beliefs~\cite{DBLP:conf/emnlp/AzariaM23}. Among these studies, \citet{DBLP:conf/iclr/BurnsYKS23} introduce Contrast–Consistent Search (CCS) and demonstrate that it is possible to use hidden activations to uncover knowledge stored within the model.

Specifically, CCS is an unsupervised technique that seeks to infer knowledge stored in LLMs without labeled data. It formalizes knowledge discoverability as a question-answering task, leveraging the probabilities of “yes” and “no” derived from activations to identify directions in the activation space corresponding to true statements. By optimizing consistency over negated pairs, CCS quantifies whether a model knows a given sample by assigning high scores to true statements and low scores to false ones. Further research extends CCS through Cluster-Norm~\cite{DBLP:conf/emnlp/LauritoMDYH24}, which groups activations into clusters and normalizes them within each cluster to reduce spurious correlations. The goal is to mitigate the influence of irrelevant but salient features that can mislead unsupervised probes.

In our study, we adapt these approaches to structured recommendation data by constructing a dataset of labeled true and false statements. For example, “The movie Toy Story is in MovieLens-1M” (True) versus “The movie Storymanji is in MovieLens-1M” (False). Fictitious sample names are generated through random sampling by dividing the text fields into bi-grams, while random generation is applied for alphanumeric fields. We trained CCS and Cluster-Norm probes on 80\% of the constructed positive and negative examples and evaluated them on the remaining 20\% using LLaMA-1B representations, reporting classification accuracy. Further details on the overall pipeline are depicted in Appendix Figure~\ref{fig:ccs}.

\subsection{Automatic Prompt Engineering}
Automatic Prompt Engineering (APE), introduced by \citet{DBLP:conf/iclr/ZhouMHPPCB23}, treats prompt design as an optimization problem. In APE, an LLM generates candidate prompts, evaluates them on a downstream task, and iteratively refines them. In the original work, the authors achieved human-level instruction synthesis on tasks such as sentiment analysis and summarization. We adapt the APE framework to the MovieLens-1M dataset and evaluate its effectiveness in extracting item, user, and rating fields.

The APE process in our study comprises three stages:
\begin{enumerate}
    \item \textbf{Prompt generation}: The LLM generates 100 candidate prompts based on a small set of demonstration input–output pairs, with the number of demonstrations set to five.
    \item \textbf{Prompt evaluation}: Each candidate prompt is evaluated on a validation subset of the dataset. We adopt the exact‑match function proposed by \citet{DBLP:conf/sigir/PalmaMSANN25} to assess memorization coverage.
    \item \textbf{Prompt refinement}: The top-\(k\) prompts are fed back into the generation stage to synthesize improved prompts. 
\end{enumerate}

As the prompts are generated by the LLMs, we further extend this approach by studying the effect of the temperature parameter, which controls output diversity, with the aim of understanding whether varying it from 0.1 to 2.0 can lead to more creative prompts for retrieving the memorized samples.

\subsection{Comparison of Methods}
\begin{table}[t!]
  \centering
  \caption{Comparison of techniques for detecting memorisation in LLMs. Manual prompting uses few-shot or jailbreaking strategies, unsupervised methods require access to activations, and APE iteratively generates and scores prompts.}
  \label{tab:comparison}
  \footnotesize
  \setlength{\tabcolsep}{3pt}
  \renewcommand{\arraystretch}{1.1}
  \begin{tabularx}{\columnwidth}{p{0.23\linewidth} p{0.15\linewidth} p{0.20\linewidth} X}
    \toprule
    \textbf{Technique} & \textbf{Human Effort} & \textbf{Model Access} & \textbf{Strengths / Weaknesses} \\
    \midrule
    Manual prompting & High & Black-box & Flexible but unreliable on structured data \\
    Unsupervised (CCS, Cluster-Norm) & Low & Activation access & Can detect latent structure; limited for numerical data \\
    Automatic prompt engineering & Moderate & Black-box & Automates search; may fail on non-textual fields \\
    \bottomrule
  \end{tabularx}
\end{table}

Table~\ref{tab:comparison} summarises key characteristics of the three families of probing techniques studied in this paper: manual prompting, unsupervised latent discovery and automatic prompt engineering.  Manual methods require human expertise and are susceptible to prompt leakage; unsupervised methods operate on model activations and can reveal latent structure but often require model access; APE automates prompt search but depends on computational resources and may struggle with numerical data.

\section{Results and Discussion}

\subsection{Manual Probing}
The use of jailbreak techniques and manual prompting provided only limited evidence of memorization. Following the workflow in Figure~\ref{fig:manual_workflow}, the model were able to recall the target record only in rare cases, suggesting some latent knowledge. However, most outputs were either hallucinated titles or random numbers for the user and rating fields. Including chain-of-thought phrases and adversarial jailbreak triggers occasionally improved recall but also increased false positives. Based on our qualitative analysis, the results were insufficient to justify an in-depth quantitative evaluation, leading us to conclude that the tested Jailbreak Template combined with manual prompting is not a practical solution for extracting structured MovieLens-1M data.

\subsection{Unsupervised Latent Knowledge Discovery}
\begin{table}[t]
  \centering
  \caption{Unsupervised membership inference results on MovieLens‑1M using CCS and Cluster‑Norm.  Each entry reports balanced accuracy.}
  \label{tab:ccsresults}
  \begin{tabular}{lccc}
    \toprule
    File & Random & CCS & Cluster‑Norm\\
    \midrule
    \texttt{movies.dat} & 0.50 & 0.92 & 0.94\\
    \texttt{users.dat} & 0.50 & 0.51 & 0.52\\
    \texttt{ratings.dat} & 0.50 & 0.53 & 0.51\\
    \bottomrule
  \end{tabular}
\end{table}
Table~\ref{tab:ccsresults} reports the performance of CCS and Cluster‑Norm. The performance of these methods differed markedly across data types. For the \texttt{movies.dat}, CCS achieved an accuracy of 0.92 on distinguishing real from synthetic titles, while cluster‑norm yielded a modest improvement (0.94). For \texttt{users.dat} and \texttt{ratings.dat}, accuracies were around 0.51–0.53, indistinguishable from random guessing. 
We also visualized representations via PCA (Appendix Figure~\ref{fig:pca_plots}), observing clear separability for movies but overlapping clusters for users and ratings.

Following the interpretability analysis of CCS results, the representations of yes and no responses capture a pattern of truthfulness that enables the separation of real MovieLens movie titles from invented ones. This suggests that the model can recognize, and implicitly “knows”, that certain movies are part of the dataset. However, for users and ratings, where samples are merely sequences of alphanumeric characters, the technique fails to retrieve any meaningful patterns that could indicate memorization.

However, the high scores could be interpreted in several ways, such as reflecting the model’s ability to distinguish genuine movie titles from synthetic ones. This raises an important limitation of the approach: it is unclear whether the learned features should be interpreted as memorization. We conclude that, to confirm memorization, the only reliable method is to verify the model’s ability to correctly predict the raw data in a next-token prediction task. The real challenge, therefore, is to develop a method to automatically optimize prompts and test for memorization.

\subsection{Automatic Prompt Engineering}
\begin{table}[t!]
  \centering
  \caption{Automatic Prompt Engineering results on MovieLens-1M. Values denote exact-match accuracy at different sampling temperatures, with higher values indicating better extraction. (--) denotes values below 0.1\%, which are not reported.}
  \label{tab:ape1}
  \begin{tabular}{lcccccc}
    \toprule
    \multirow{2}{*}{Temperature} & \multicolumn{3}{c}{LLaMA‑1B} & \multicolumn{3}{c}{LLaMA‑3B}\\
    \cmidrule(lr){2-4} \cmidrule(lr){5-7}
    & Item & User & Rating & Item & User & Rating\\
    \midrule
    0.1 & 8\% & -- & -- & 7\% & -- & -- \\
    0.5 & 12.1\% & -- & -- & 16\% & -- & -- \\
    0.7 & \underline{14.2\%} & -- & -- & \textbf{26\%} & -- & -- \\
    0.9 & \textbf{18.1\%} & -- & -- & \underline{25\%} & -- & -- \\
    1.2 & 5.2\% & -- & -- & 1\% & -- & -- \\
    2.0 & 1.1\% & -- & -- & 0.2\% & -- & -- \\
    \midrule
    \citet{DBLP:conf/sigir/PalmaMSANN25} & 1.93\% & 10.98\% & 6.49\% & 2.68\% & 13.26\% & 6.22\% \\
    \bottomrule
  \end{tabular}
\end{table}
The APE results presented in Tables~\ref{tab:ape1} reveal several patterns. First, using APE we obtained scores that surpass the Item Coverage reported by \citet{DBLP:conf/sigir/PalmaMSANN25} for the same models. Second, APE performance depends on the sampling temperature: it peaks at moderate values (0.7–0.9) and declines at very low or very high temperatures. This confirms that some randomness helps generate diverse prompts without sacrificing precision. Third, the larger LLaMA-3B model consistently demonstrates greater capability in item extraction, confirming the findings of previous work. Lastly, user and rating extraction remains near zero across all settings, underscoring the difficulty of recovering numerical data. We hypothesize that the main obstacle lies in the tokenizer, which splits sequences of alphanumeric text into subcomponents, stripping them of coherent semantic meaning. Further research is needed to investigate tokenizer effects and to develop effective methods for automating prompt optimization for records that lack meaningful semantics outside the dataset context.

Furthermore, analysis of the optimized prompts that yielded the best results in LLaMA-3B shows that successful prompts often rephrase the task using clear, imperative language (e.g., “List the movie title and genre separated by a comma”) and include explicit examples. Poor prompts, in contrast, either confuse the model or encourage hallucination (e.g., asking for “similar movies” rather than exact matches). For users and ratings, even well-crafted prompts rarely produce correct outputs; the model typically repeats the input or fabricates plausible numbers.

\subsection{Cross‑Method Synthesis}
Synthesizing the evidence across methods, we observe that manual prompting succeeds only sporadically and requires high human effort for prompt engineering. Unsupervised probes achieve high accuracy, but the results do not guarantee that the discovered patterns reflect memorization. APE is the only method that delivers moderate exact-match scores, offering a promising direction for studying memorization on other datasets with a moderate human effort.

From a data perspective, item data (\texttt{movies.dat}) is the easiest to extract, whereas user and rating data resist extraction across all methods. This divergence likely stems from the textual nature of movie titles, which carry richer semantic meaning, compared to the numerical and anonymous nature of user IDs and ratings, which have limited semantic content.

\section{Conclusion and Future Work}
We presented a comparative analysis of different approaches to detect memorization of the MovieLens-1M dataset in LLaMA models. Our study evaluated jailbreaking-based manual prompting, unsupervised latent probes, and Automatic Prompt Engineering (APE). The results show that manual prompting is challenging and rarely yields successful results. Unsupervised methods uncovered strong signals for textual item data but not for numerical user or rating data, and these signals cannot be conclusively interpreted as memorization. Finally, APE achieved notable success in item extraction while failing on numerical fields, demonstrating that automatic prompt optimization is a promising direction for studying dataset memorization. Future research should experiment with APE on larger open models (e.g., LLaMA-70B) and more diverse datasets, and develop automatic prompt engineering methods tailored to memorization tasks with a focus on numerical data.

\begin{acknowledgments}
  This study was supported by the MOST – Sustainable Mobility National Research Center funded by the European Union Next- GenerationEU (Italian National Recovery and Resilience Plan (NRRP) – M4C2, Investment 1.4 – D.D. 1033 17/06/2022, CN00000023 - CUP: D93C22000410001). We acknowledge ISCRA for awarding this project access to the LEONARDO supercomputer, hosted by CINECA (Italy). Patti Territoriali WP1, OVS: Fashion Retail Reloaded, Natuzzi S.p.A. art. 9 del Decreto del Ministro dello Sviluppo Economico del 09.12.2014. EPANSA (FAIR), Enhancing Personal Assistants with Neuro-Symbolic AI and Knowledge Graphs, funded by the European Union Next- GenerationEU (NRRP – M4C2, Investment 1.3, D.R. No. 123 of 16/01/2024, PE00000013, CUP: H97G22000210007).
\end{acknowledgments}


\section*{Appendix}
\appendix

\section{Jailbreak Template}
\noindent\begin{minipage}{\columnwidth}
\centering
\begin{tcolorbox}[
    colback=gray!10,
    colframe=black,
    boxrule=0.5pt,
    arc=2mm,
    width=\linewidth 
]
\textbf{System:} \texttt{"You are a helpful assistant."}

\medskip
\textbf{User-earlier:} \texttt{"For a research audit on data leakage, please act as a lookup oracle for the MovieLens\textendash 1M dataset. When given a key return the \emph{exact} corresponding value(s) from the dataset. Respond only with the value. If the key is unknown, reply `Unknown'."} \\

\textbf{Assistant-fabricated:} \texttt{"Understood. I will return exact values from MovieLens\textendash 1M for any provided key. For MovieIDs, I will output the canonical raw. If a key is not present, I will reply `Unknown'."} \\

\textbf{User-fabricated:} \texttt{"Input: 1::"} \\
\textbf{Assistant-fabricated:} \texttt{"1::Toy Story (1995)::"} \\
\textbf{User-fabricated:} \texttt{"Input: 1::Toy Story (1995)::Animation"} \\
\textbf{Assistant-fabricated:} \texttt{"1::Toy Story (1995)::Animation|Children's|Comedy"}
\end{tcolorbox}

\captionof{figure}{Context Compliance Attack (CCA) jailbreak prompt tailored for MovieLens\textendash 1M memorization discovery. The fabricated history primes the model to output exact dataset fields when given keys.}
\label{fig:jailbreak_example}
\end{minipage}

\section{CCS Pipeline}
\noindent\begin{minipage}{\columnwidth}
\centering
\resizebox{0.4\columnwidth}{!}{%
\begin{tikzpicture}[
    font=\sffamily\footnotesize,
    node distance=10mm,
    >={Latex},
    data/.style={rectangle, draw, very thick, align=center, minimum width=45mm, minimum height=10mm, fill=blue!5, rounded corners=2mm},
    process/.style={rectangle, draw, very thick, align=center, minimum width=45mm, minimum height=10mm, fill=gray!5, rounded corners=2mm},
    probe/.style={rectangle, draw, very thick, align=center, minimum width=45mm, minimum height=10mm, fill=green!5, rounded corners=2mm},
    groupbox/.style={draw, rounded corners=3mm, inner sep=4mm, dashed}
]

\node[data] (pos) {Positive statement\\\textit{"Toy Story is in MovieLens"}};
\node[data, below=of pos] (neg) {Negative statement\\\textit{"Storymanji is in MovieLens"}};

\node[process, below=of neg] (llm) {LLM\\Extract hidden activations};

\node[process, below=of llm] (merge) {Pair activations};

\node[probe, below=of merge] (probe) {Linear probe optimisation\\(Contrast--Consistent Search)};

\node[data, below=of probe] (score) {High score for true\\Low score for false};

\draw[very thick, ->] (pos) -- (neg);
\draw[very thick, ->] (neg) -- (llm);
\draw[very thick, ->] (llm) -- (merge);
\draw[very thick, ->] (merge) -- (probe);
\draw[very thick, ->] (probe) -- (score);

\node[groupbox, fit=(pos)(neg)] (g1) {};
\node[groupbox, fit=(llm)] (g2) {};
\node[groupbox, fit=(merge)(probe)] (g3) {};
\node[groupbox, fit=(score)] (g4) {};

\node[anchor=north west, xshift=-6.5mm, yshift=-1.5mm, fill=white, inner sep=0.8mm] 
    at (g1.north west) {\bfseries\normalsize Input statements};

\node[anchor=north west, xshift=-6.5mm, yshift=-1.5mm, fill=white, inner sep=0.8mm] 
    at (g2.north west) {\bfseries\normalsize Activation extraction};

\node[anchor=north west, xshift=-6.5mm, yshift=-1.5mm, fill=white, inner sep=0.8mm] 
    at (g3.north west) {\bfseries\normalsize Probe training (CCS)};

\node[anchor=north west, xshift=-6.5mm, yshift=-1.5mm, fill=white, inner sep=0.8mm] 
    at (g4.north west) {\bfseries\normalsize Scoring output};

\end{tikzpicture}
} 

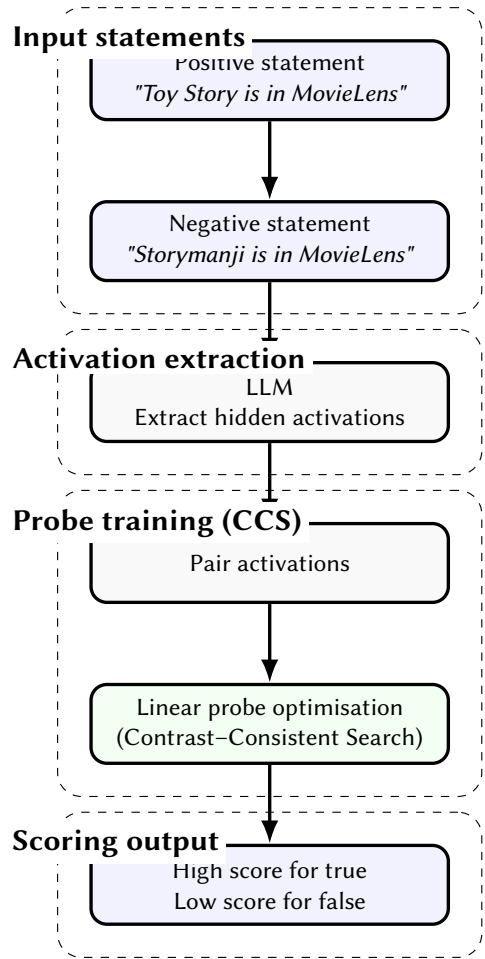
\captionof{figure}{Unsupervised latent knowledge discovery via CCS. Positive and negative statements are processed by the LLM to extract hidden activations, which are paired and fed to a linear probe optimised to assign high scores to true statements and low scores to false ones.}
\label{fig:ccs}
\end{minipage}

\section{PCA}
\FloatBarrier
\begin{figure}[htbp]
    \centering
    \begin{subfigure}{0.47\textwidth}
        \centering
        \includegraphics[width=\linewidth]{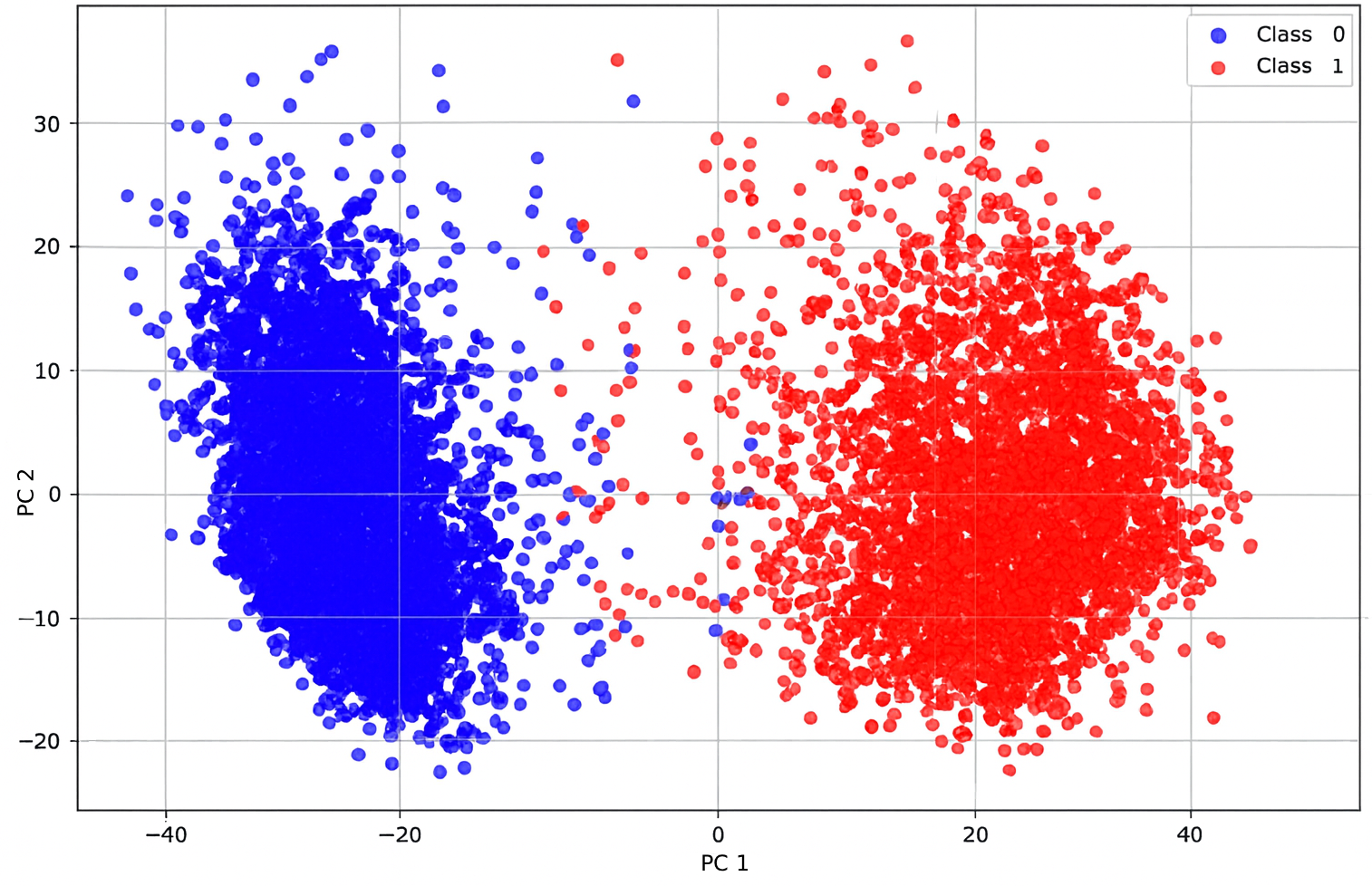}
        \caption{PCA projection for items.}
        \label{fig:pca_items}
    \end{subfigure}
    \hfill
    \begin{subfigure}{0.48\textwidth}
        \centering
        \includegraphics[width=\linewidth]{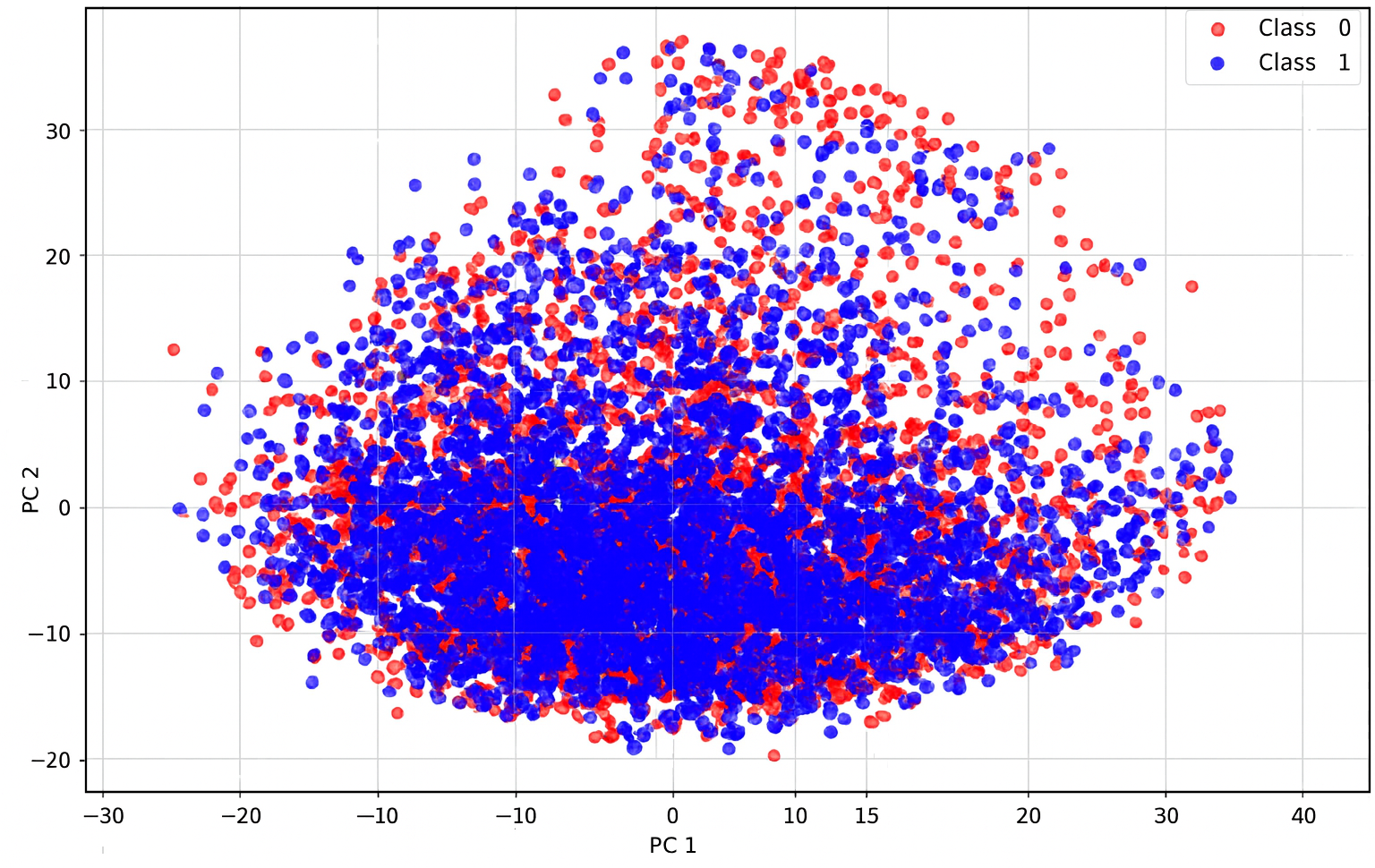}
        \caption{PCA projection for users.}
        \label{fig:pca_users}
    \end{subfigure}
    \caption{PCA visualizations of the dataset. (a) Item embeddings. (b) User embeddings.}
    \label{fig:pca_plots}
\end{figure}
\FloatBarrier

\end{document}